\documentclass[prl,aps,reprint,noshowpacs,superscriptaddress,floatfix,letterpaper,longbibliography]{revtex4-2}
\usepackage{amsmath,amssymb,amsbsy,amsfonts,amsthm,bbm,bm,mathtools,mathrsfs}
\usepackage{color}
\usepackage{physics}
\usepackage{xfrac}
\usepackage[dvipsnames]{xcolor}
\definecolor{LapisLazuli}{RGB}{47, 102, 169}
\usepackage[colorlinks=true,citecolor=LapisLazuli,linkcolor=LapisLazuli,urlcolor=LapisLazuli]{hyperref}
\usepackage{empheq}
\usepackage{stackengine}
\usepackage{relsize}
\usepackage[inline]{enumitem}
\usepackage[normalem]{ulem}
\usepackage{comment}
\usepackage{import}
\usepackage[english]{babel}
\usepackage{lipsum} 
\usepackage{graphicx}
\usepackage{soul}

\newcommand{\jrg}[1]{\textcolor{black}{#1}}

\date{\today}

\begin{document}

\title{Dissipation rates from experimental uncertainty}

\author{Aishani Ghosal}
\author{Jason~R.~Green}
\email[]{jason.green@umb.edu}
\affiliation{Department of Chemistry,\
	University of Massachusetts Boston,\
	Boston, MA 02125 USA
}
\affiliation{Department of Physics,\
	University of Massachusetts Boston,\
	Boston, MA 02125 USA
}
\date{\today}

\begin{abstract}	
\jrg{
Active matter and driven systems exhibit statistical fluctuations in density and particle positions, providing an indirect indicator of dissipation across multiple length and time scales.
Here, we quantitatively relate these measurable fluctuations to a thermodynamic speed limit that constrains the rates of heat and entropy production in nonequilibrium processes.
By reparametrizing the speed limit, we show how to infer heat and entropy production rates from directly observable or controllable quantities.
This approach can use available experimental data and avoid the need for analytically solvable microscopic models or full time-dependent probability distributions.
The heat rate we predict agrees with experimental measurements for a Brownian particle and a microtubule active gel, which validates the approach and suggests potential for the design of experiments.
}

\end{abstract}

\maketitle

\textit{Introduction.--} Dissipation rates determine the efficiency of biological processes~\cite{FakhriPRL, Fakhri2016} and are now an important feature in designing active synthetic materials that function~\cite{duclos2020topological,kay2007synthetic,rossumDissipativeOutofequilibriumAssembly2017,ragazzonEnergyConsumptionChemical2018,dasChemicallyFueledSelfAssembly2021,del2022dissipative,barpuzary2023waste,shklyaev2024interlinking}.
For example, active cytoskeletal materials \jrg{sustain self-organized flows by consuming chemical energy and dissipating} heat at picowatt rates~\cite{ActiveGel}.
It recently became possible to measure these rates with advances in picocalorimetry~\cite{bae2021micromachined} and to directly observe active materials in the liquid phase with high-resolution electron microscopy~\cite{rizviCloseLookMolecular2021}. 
These techniques are able to resolve dissipative degrees of freedom and beginning to provide a better understanding of how chemical energy propagates up length and timescales. 
\jrg{However, these measurements are often interpreted with empirical models of the kinetics, which can have large uncertainties, missing parameters, and modeling assumptions that may be unjustified for active materials (e.g., well-mixed chemical kinetics, modeling only rate limiting reactions)~\cite{nicolaouMissingLinksSource2020}.}
\jrg{Model-free approaches to predict dissipation rates} from available experimental data could guide this experimentation on functional materials and their design~\cite{van2017dynamic,moulin2020molecular,zhang2023electric}.

Surveying stochastic thermodynamics~\cite{pelitiStochasticThermodynamics2021}, the thermodynamic speed limit~\cite{Nicholson:2020} set by the Fisher information~\cite{fisher1925theory,kim2021information,frieden2004science} $I_{F}=\tau^{-2}$ has potential for estimating dissipation rates. 
\jrg{While not yet experimentally tested, t}his single timescale upper bounds the rate of heat, entropy production, dissipated work, chemical work~\cite{Nicholson:2020,nicholson2021thermodynamic,aghion2023relations,aghion2023thermodynamic}. 
For example, a system that dissipates energy as heat with a rate $\dot{Q}$~\cite{oono1998steady} 
and is subject to energy fluctuations with a standard deviation $\Delta\epsilon$ has the speed limit~\cite{Nicholson:2020},
\begin{equation} 
	\tau_{Q}^{-1} := \frac{|\dot{Q}|}{\Delta\epsilon} \leq \sqrt{I_F} =: \tau^{-1},
	\label{eq:tiur} 
\end{equation}
even if work is done on or by the system~\footnote{If no work is done on or by the system, then a simpler result holds~\cite{ito2020stochastic}.}. 
Speed limits on dissipation rates are now known in quantum~\cite{DeffnerPRL2013,DeffnerPRL2020,DeffnerPRR2020}, classical deterministic~\cite{Das2023, Das2024}, and stochastic~\cite{SLclassicalstosys, nicholsonNonequilibriumUncertaintyPrinciple2018,hasegawaUncertaintyRelationsStochastic2019, Nicholson:2020} dynamics. 
Alternative approaches to infer dissipation, such as the \jrg{dissipation-time uncertainty relation~\cite{falasco2020dissipation}} or the thermodynamic uncertainty relation~\cite{barato2015thermodynamic,GingrichFPT2017,li2019quantifying,horowitz2020thermodynamic,manikandan2020inferring,Ostubu2020PRE,van2020entropy,pal2021thermodynamic} and its extensions~\cite{dechant2021improving,pietzonka2023thermodynamic} estimate dissipation through the statistics of currents~\cite{lyu2024entropy}. 
Others use optimization methods and transition rates~\cite{skinner2021estimating,nitzan2022universal}, waiting times~\cite{skinner2021estimating, van2022thermodynamic}, \jrg{variances~\cite{di2024variance,di2024variance1}, partial information, and coarse-graining ~\cite{PhysRevX.12.041026, blom2024milestoning}}.
However, these approaches give lower bounds on entropy production rates that are not necessarily explicit functions of easily measurable observables or estimates of the energy dissipated as heat.
By contrast, estimates of the Fisher information in Eq.~\eqref{eq:tiur} could give upper bounds on the maximum rate of dissipation of entropy and heat.
Eq.~\eqref{eq:tiur} is also unique in that it can be combined with the quantum time-energy uncertainty relation~\cite{mandelstamUncertaintyRelationEnergy1991} to bound the dynamical observables of open quantum systems~\cite{garcia-pintosUnifyingQuantumClassical2022}.

In this Letter, we establish a method for predicting the rates of nonequilibrium observables from experimental data through the thermodynamic speed limit in Eq.~\eqref{eq:tiur}.
The Fisher information enables us to use measurements of uncertainty as input. 
Both classical and quantum Fisher information are part of the statistical design of experiments on systems ranging from chemical reactions and biological populations to dark matter~\cite{jung2021optimal}; by measuring experimental errors, one can use the Fisher information (matrix) to predict \textit{a priori} the minimum error of any measured quantity~\cite{berengut2006statistics}.
In this context, the Fisher information predicts the minimum error in an observable before the measurement, which can then be used to assess the measurement sensitivity to control variables, minimize errors, and improve the precision.
\jrg{Here, we establish a similar framework for inferring dissipation rates, transforming the coordinates of the Fisher information so that we can use available measurements of positions, concentrations, and control variables}~\cite{ActiveGel}. 
Our predictions confirm experimentally measured dissipation rates in several examples, including an active cytoskeleton materials composed of kinesin motors and microtubules~\cite{ActiveGel} measured with picocalorimetry~\cite{bae2021micromachined}.

\begin{figure}
  \includegraphics[width=1.00\columnwidth]{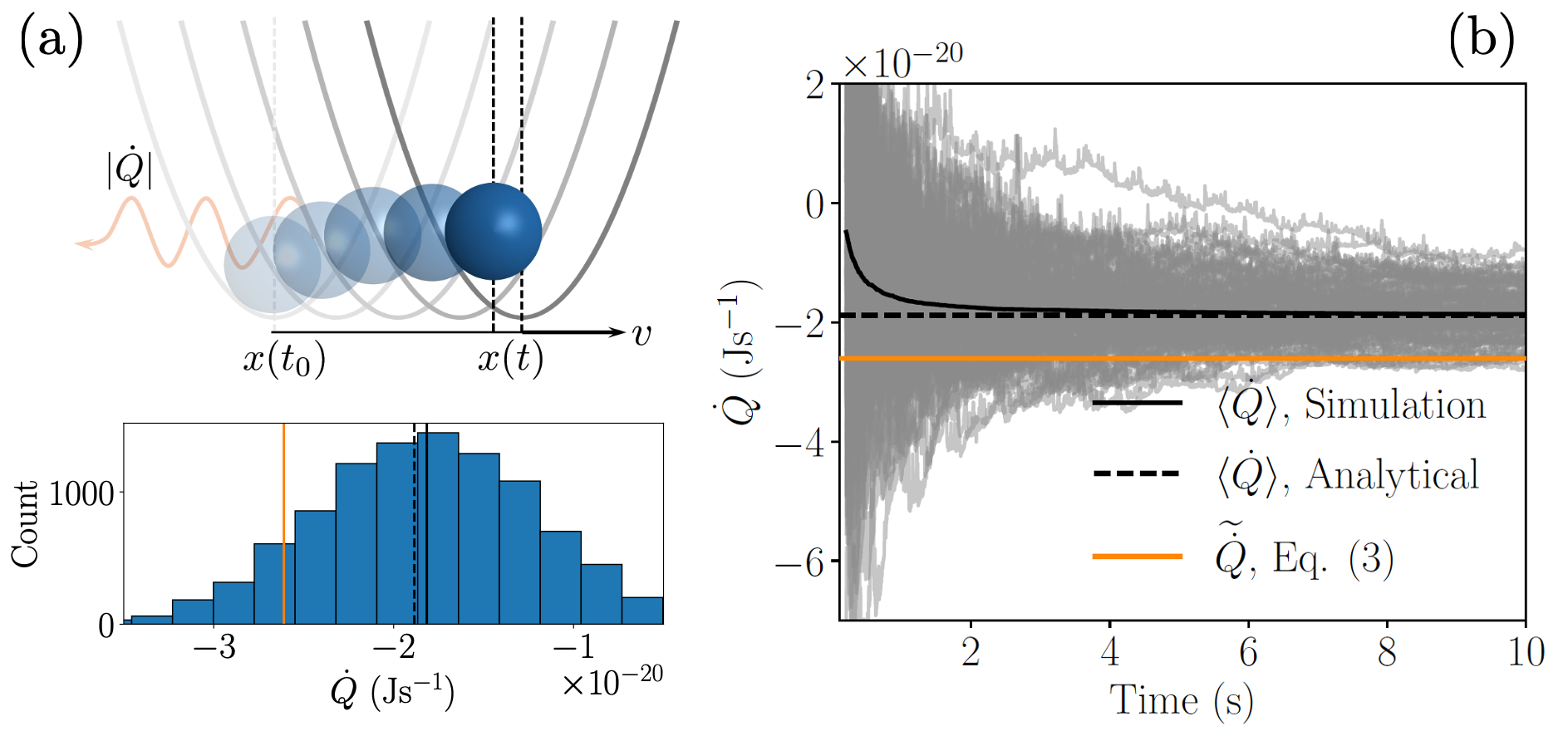}
  \caption{\label{fig:2}(a) A Brownian particle dissipates energy as heat when
pulled through a viscous medium. At $t=t_{0}$, the particle is at the trap
center $x(t_0)$. Translating the trap through space forces the average particle
position to be higher on the potential. During the process, heat is dissipated at a
rate $|\dot{Q}|$ into the environment. (b) Histogram of the heat dissipated to the
environment from a simulation of the overdamped Langevin dynamics for 10,000 noise
realizations at $t=4$\,s (SM Sec.~2). \jrg{The housekeeping} heat rate from Eq.~\eqref{eq:heatexpmovingtrap} (orange) agrees with the sample mean from the simulations (black solid line)
and the analytical mean (black dotted line).}
\end{figure}

\textit{Predicting dissipated heat by transforming the speed limit.--} To illustrate the approach, first consider the net rate of heat generation $|\dot{Q}|$ when dragging an optically trapped colloidal particle at a speed $|v|$ through a viscous medium at a temperature $T$, \jrg{Fig.~\ref{fig:2}.
The position fluctuates with a standard deviation $\sigma_{x}$ that converges to a finite value at nonequilibrium steady state. 
We consider the case in which the true position distribution is unknown but one can estimate statistical parameters by repeated measurements.} 
In this case, repeatedly measuring the spatial location of the particle using an estimator $\hat{x}$ will lead to an empirical standard deviation $\Delta \hat{x} = \sigma_x \pm |\varepsilon|$ with measurement error $\varepsilon$. 
\jrg{
We denote estimators of directly measured variables \unexpanded{$\hat{\theta}$} to distinguish them from the true values \unexpanded{$\theta$} and those predicted \unexpanded{$\tilde{\theta}$}.}

At first glance, determining the speed limit set by the Fisher information seems to require the probability distribution $\rho(x,t)$ from experiments or a \jrg{microscopic model for the dynamics of the particle}. 
The measured positions $x_i$ depend on the \jrg{unknown} distribution $\rho(x,t)$ \jrg{moving through space} at an \jrg{unknown} intrinsic rate $\hat{r} = -\partial_t\ln \rho(x,t)$ \jrg{-- a nonzero quantity in the lab frame at nonequilibrium steady state}.
However, because the Fisher information is a variance $\Delta \hat{r}^{2}$, we can change its coordinates and effectively propagate the error from the uncertainty in the particle position (Supplementary Material, SM Sec.~1).
\jrg{Through this coordinate transformation, a direct measurement of $\Delta \hat{x}$ can serve as an indirect measurement $\tilde{\tau}^{-1}=\Delta\hat{r}$ of the true speed limit in Eq.~\eqref{eq:tiur}.}

\jrg{To relate the speed limit to data, we recognize that the fluctuations in position $\Delta \hat{x}$ propagate~\cite{gauss1823theoria} to $\Delta\hat{r}$.} 
Since the estimator $\hat{r}$ is a function $\hat{r} = f(\hat{x})$, the speed limit is
\begin{equation}
	\label{eq:EP}
	\tilde{\tau}^{-1}:=\Delta\hat{r} = \frac{\Delta \hat{x}}{|\partial_{f}\hat{x}|} + \mathcal{O}(\Delta \hat{x}^2).
\end{equation}
Rearranging to $\Delta \hat x^2 = (\partial_f\hat{x})\tilde{\tau}^{-2}(\partial_f\hat{x})$, we can also recognize \jrg{that the uncertainty in position is a coordinate transformation of the Fisher information.  
This transformation is a special case of our main result: it relates measurable uncertainties to the speed limit, 
which can enables estimates of the speed limit without a dynamical model for the probability distribution.
}

\jrg{When the dynamics are unknown, we can find the derivatives in Eq.~\eqref{eq:EP} using regression techniques. 
By hypothesizing the relationship between the rate $\hat{r}=f(\hat{x})$~\cite{nicholson2021thermodynamic} and chosen observable $\hat{x}$, we can minimize the mean squared error to find the optimal parameters and the derivatives~\cite{barlow1993statistics}.} 
For example, hypothesizing a linear relationship $f(\hat{x})=a+b\hat{x}$, the optimal slope is $b_{\text{opt}}=\partial_{\hat{x}}f=v/\Delta\hat{x}^2$ (SM Sec.~2).
Using the result in Eq.~\eqref{eq:EP}, the speed limit $\tilde{\tau}^{-1}=|v|/\Delta\hat{x}$ 
is in terms of the pulling speed and sample variance $\Delta\hat{x}$.

\jrg{With direct measurements or estimates of the position fluctuations, we can also use the speed limit to predict the heat rate \jrg{(SM Sec.~3)} and other nonequilibrium observables.} 
Notice that we can isolate the heat rate in Eq.~\eqref{eq:tiur}: $|\dot{Q}|=\tau_{Q}^{-1}\Delta\epsilon$. 
If we similarly find the measured heat rate $\widetilde{\dot{Q}}$ using Eq.~\eqref{eq:EP} and estimating $\widetilde{\tau}_{Q}^{-1}\Delta\hat{\epsilon}\approx\tilde{\tau}^{-1}\Delta\hat{\epsilon}$.
For the pulled particle, the heat rate at nonequilibrium steady state
\begin{equation}
\label{eq:heatexpmovingtrap}
|\widetilde{\dot{Q}}| = \tilde{\tau}_{Q}^{-1} \Delta\hat\epsilon \approx |v|\Delta\hat\epsilon/\Delta \hat{x}, 
\end{equation}
is in terms of known quantities from the experimental setup: the pulling speed and the uncertainties in the particle position and energy. 
\jrg{In the lab frame, this heat rate is the housekeeping contribution that maintains the nonequilibrium steady state~\cite{oono1998steady}.}

\jrg{From Eq.~\eqref{eq:heatexpmovingtrap}, we can predict the heat rate using available experimental parameters. Take a} particle of radius $1\,\mu$m being pulled at a speed~\cite{Imparato} $|v|=1$\,$\mu$m\,s$^{-1}$ through water. 
\jrg{If the standard deviations in Eq.~\eqref{eq:heatexpmovingtrap} are obtained from measurements, they can be directly used to calculate the heat rate without additional assumptions.}
However, \jrg{provided the surrounding medium is at thermal equilibrium}, the standard deviation in position is known~\cite{ritort2006single} to be $\Delta \hat{x} \approx \sigma_{{x}}=\sqrt{k_{B}T/k_f} = 78.32$\,nm (errors can be of order $\pm 2$\,nm~\cite{Imparato}), and the standard deviation in energy is $\Delta \hat{\epsilon} \approx k_{B}T/2$.
We take the trap to be harmonic with a force constant $k_f=6.67 \times 10^{-7}$\,N\,m$^{-1}$ and the surrounding medium to be water with a viscosity $\eta=10^{-3}$\,Pa\,s at 296.5\,K~\footnote{With these numerical values of radius and viscosity, the inverse friction coefficient is $\gamma = 5.30 \times 10^{7}$\,m\,N$^{-1}$\,s$^{-1}$.}.
Using only these experimental values, the heat rate from Eq.~\eqref{eq:heatexpmovingtrap} is $|\widetilde{\dot{Q}}|=|v|\sqrt{k_f/\beta}=2.61 \times 10^{-20}$\,W. 

\jrg{To confirm this value, we chose physical models for the dynamics of the particle at the same experimental conditions. 
We numerically simulated a particle undergoing Brownian dynamics and, over time, measured the particle position relative to the trap center in the laboratory frame~\cite{speck2008role}~\footnote{In the comoving frame~\cite{movingtrapposdis}, the position is initially delta function distributed $\rho(x-vt_0)=\delta(x-x(t_0))$ and eventually evolves as $\rho(x-vt) = e^{-|x-vt|^2/2\sigma_{x}^{2}}/\sqrt{2 \pi \sigma_{x}^{2}}$ at steady state.}}. 
\jrg{As shown in Fig.~\ref{fig:2}, the housekeeping heat rate from Eq.~\eqref{eq:heatexpmovingtrap} agrees well with the average over noise realizations from our numerical simulations $1.85 \times 10^{-20}$\,W.
It also agrees well with an analytical prediction~\cite{Imparato} for the heat rate $|\langle \dot{Q} \rangle|=v^2/\gamma = 1.885 \times 10^{-20}$\,W \jrg{(SM Sec~2)}.}

\begin{figure}
\includegraphics[width=0.85\columnwidth]{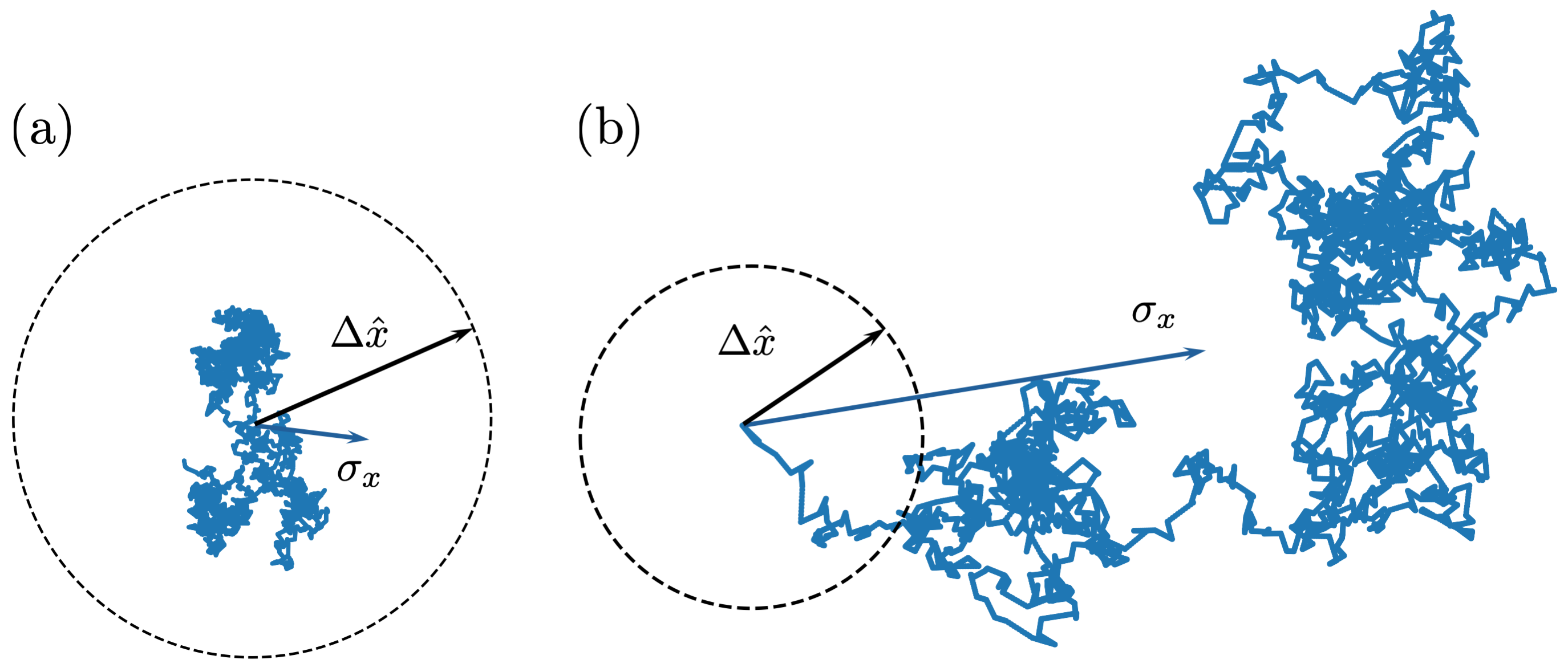}
\caption{\label{fig:1}
\jrg{Illustration showing the spatial resolution of experimental measurements compared to the intrinsic fluctuations of a particle undergoing Brownian motion.
(a) A lower resolution (higher uncertainty, $\Delta\hat{x}$) than required to measure position fluctuations $\sigma_x$ of a Brownian particle leads to an upper bound on the speed limit, $\tilde{\tau}^{-1}\leq \tau^{-1}$.} (b) A higher resolution (lower uncertainty) than required to measure position fluctuations leads to a lower bound, $\tilde{\tau}^{-1}\geq \tau^{-1}$.
}
\end{figure} 

\textit{Thermodynamic speed limits from experimental uncertainty.--}
As this example highlights, the coordinate transformation of the Fisher information, together with a regression hypothesis for the relationship between $\hat r$ and $\hat x$, can bypass the need for the full probability distribution of a dynamical model. 
This observation motivated us to derive a more general framework for measuring this speed limit by transforming the Fisher information into the variance of other measurable quantities. 

Take $\boldsymbol{X}$ to be a vector of quantities that are measurable by experiment.
Because of their statistical fluctuations, we can model these quantities as random variables with means $\boldsymbol{\mu}$ and covariances $(\boldsymbol{C}_{\boldsymbol{X}})_{ij} = \operatorname{cov}\left(X_i, X_j\right)$. 
But, imagine the system evolves under a dynamics described by random variables $\boldsymbol{Y}$ that are functions $\boldsymbol{Y}=\boldsymbol{f}(\boldsymbol{X})$. 
In this case, we can interpret the covariance matrix $(\boldsymbol{C}_Y)_{kl} = \operatorname{cov}\left(f_k, f_l\right)$ as the Fisher matrix that becomes the speed limit.

As in the pulled particle example, we recognize that quantities in the speed limit are (co)variances and use this fact to transform the speed limit.
To start, we Taylor expand the $m$ functions of $\boldsymbol{f}$ in terms of $n$ measured quantities $\boldsymbol{X}$
\begin{equation}
	f_i(\boldsymbol{X})=f_i(\boldsymbol{\mu})+\grad f_i^\top|_{\boldsymbol{X}=\boldsymbol{\mu}}(\boldsymbol{X}-\boldsymbol{\mu})+\mathcal{E}
\end{equation}
with the gradient $\grad f_i=(\partial_{X_1} f_i,\partial_{X_2} f_i,\ldots)^\top$ and error $\mathcal{E}$.
Taking the error to be uncorrelated with the inputs $X_i$, the covariances 
between elements of $\boldsymbol{X}$ propagate to the covariances of $\boldsymbol{f}$ 
\begin{equation}
\label{eq:EPgeneral}
\boldsymbol{C}_Y=\grad \boldsymbol{f}^\top\boldsymbol{C}_X\grad \boldsymbol{f}+\sigma^2_{\mathcal{E}}
\end{equation}
under the action of the Jacobian $\grad \boldsymbol{f}$.
We can interpret this relationship as a generalization of Gauss's error propagation law~\cite{gauss1823theoria,bevingtonDataReductionError2002} or as the coordinate transformation of the Fisher matrix from $\boldsymbol{X}$ to $\boldsymbol{Y}$~\cite{lehmannTheoryPointEstimation1998}. 
Even if the measurable variables are correlated with covariance matrix $\boldsymbol{C}_X$, one can predict the variance of several $\boldsymbol{f}(\boldsymbol{X})$ or a single function $Y=f(\boldsymbol{X})$~\footnote{Standard propagation of error is a simple approximation of the distribution $\rho_Y(y)$ of $Y$ with only the first two moments, $\mu_Y$ and $\sigma_Y^2$:
Gauss's error propagation law~\cite{gauss1823theoria} predicts the variance 
$\sigma_{Y}^{2}= \sigma_{X}^{2} (\partial_{X}f|_{X=\mu_{X}})^{2}$ and the Taylor expansion $Y=f(\mu_X)+\partial_X f|_{X=\mu_X}(X-\mu_X)$ predicts the mean $\mu_Y=f(\mu_X)$. 
It is a special case in which the function $f$ is linear, making the error $\varepsilon=0$ and $\sigma^2_{\varepsilon}=0$.}.

With Eq.~\eqref{eq:EPgeneral}, we can transform the scalar Fisher information that sets the speed limit. 
When $Y=\hat r$, the Fisher information transforms as
\begin{equation}
\label{eq:mvspeed}
\tilde{\tau}^{-2} = \Delta \hat{r}^2 = \grad f^\top(\boldsymbol{X})\, \boldsymbol{C}_{\boldsymbol{X}}\,\grad f(\boldsymbol{X}).
\end{equation}
\jrg{This form of the speed limit is our main result: it transforms the directly measurable correlations between $\boldsymbol{X}$ to the speed limit $\tilde{\tau}^{-1}$.} 
For a single measurable quantity the Fisher information transforms from $I_{F}(\theta) = (\partial_{\theta}f(\theta))^{2} I_{F}(f(\theta))$ for a parameter, $I_{F}(\theta)$, to a function of that parameter, $I_{F}(f(\theta)) = \Delta \theta^{2}$. 
Equation~\eqref{eq:mvspeed} becomes Eq.~\eqref{eq:heatexpmovingtrap} by recognizing that 
the Jacobian from $\theta = \hat x$ to $f(\theta) = \hat{r}$ is $|\partial_{\hat x}\hat{r}| = |d_{t}{\hat x}|/\Delta\hat{x}^{2} = |v|/\Delta \hat{x}^{2}$. 

\textit{Inferring the speed limit from data.--}All together, this transformation of the speed limit allows us to choose the input (co)variances in $\boldsymbol{C}_X$ based on experimental convenience and the data that is available. 
As with any estimate, these choices must be made judiciously. 
(i) The observables, such as the heat rate $\dot{Q}=-\operatorname{cov}(\hat{r},\epsilon)$, are expressible as a covariance with the rate $\hat{r}$ (e.g., dissipated work, chemical work, entropy production, entropy flow).
\jrg{The input variables are also those that are dissipative and have a finite variance; measuring fluctuations in variables that are uncorrelated with dissipative degrees of freedom would give less accurate estimates of the speed limit (SM Sec.~3).} 

(ii) \jrg{When the distribution is Gaussian,} the dragged colloidal particle is an example of the case in which $Y=\hat r$ is a linear function of $\boldsymbol{X}\to\hat{X}$.
However, the dynamics may not be Brownian, the distribution of $\hat{X}$ may not be Gaussian, or the distribution may not be known analytically.
\jrg{Under these circumstances, the true relationship between $\hat{r}$ and $\hat{X}$ may be nonlinear. 
While a linear regression hypothesis would neglect higher order terms, a linear hypothesis will still perform well when the parameter $b$ is constant over several $\Delta \hat{X}$~\cite{barlow1993statistics}. 
In the nonequilibrium steady state examples here, this is a good approximation because $b=\partial_{X} \hat{r}$ is a nonzero constant (i.e., the trap velocity and the time evolution of ATP concentration for the active gel below).} 

\jrg{Even if the true relationship is nonlinear, one can still minimize the error, determine the parameters, and make an optimal prediction $\tilde{\tau}^{-1}$ of the true speed limit $\tau^{-1}$ with experimental data.}
Linear functions $\hat r=a+b\hat{X}$ also have the advantage that they saturate the inequality $\tilde{\tau}_Q^{-1}=\tilde{\tau}^{-1}$~\cite{nicholson2021thermodynamic} with an optimal slope $b_{\textrm{opt}}=\operatorname{cov}(\hat{X},\hat{r})/\Delta \hat{X}^2=\dot{\mathcal{X}}/\Delta \hat{X}^2$~\cite{nicholson2021thermodynamic}.
So, estimates of the speed limit are possible with linear models (avoiding the nonequilibrium probability distribution over time) when the rate $\dot{\mathcal{X}}$ and variance $\Delta \hat{X}^2$ are available. 

\jrg{(iii) Another important consideration in estimating the speed limit from data is when the measurement overestimates the true speed limit $\tau_O^{-1}\leq \tau^{-1}\leq \tilde{\tau}^{-1}$.
For an observable $O$, we found the transformed speed limit $\tilde{\tau}^{-1}$ upper bounds the heat rate, entropy production rate, and dissipated work when the empirical standard deviation $\Delta \hat{O}$ is larger than the true parameter $\sigma_{O}$. 
For example, in the pulled particle example $O=\hat{x}$, the ratio of the true and error propagated speed limits is
\begin{equation}
\label{eq:ratio}
\frac{\tau^{-1}}{\tilde{\tau}^{-1}} = \frac{\tau^{-1}}{\tilde{\tau}_{Q}^{-1}}= \frac{\sigma_{x}}{\Delta \hat{x}}.
\end{equation}
\jrg{This equality holds for systems with a linear or quadratic relation between $\hat{r}$ and the experimentally accessible variable $\hat{x}$ (SM Sec.~4-5).} 
When $\Delta \hat{x} > \sigma_{x}$, the empirical speed limit will overestimate the true speed limit $\tau^{-1} < \tilde{\tau}^{-1}$, Fig.~\ref{fig:1}.
The predicted speed limit is exact when there is no measurement error $\varepsilon$ and \jrg{$\Delta\hat{x}=\sigma_x$}. 
For the pulled particle, the ratio of the Fisher informations satisfies this equality:
The true Fisher information for the dragged particle is $I_F=\tau^{-2}=v^2 \sigma_{x}^2$ and the Fisher information propagated from the error in $\hat{x}$ is $\tilde{I}_F = \tilde{\tau}^{-1}=v^2 \Delta \hat{x}^2$ (SM Sec.2).}
It also follows that the measurement error determines whether predictions of $|\widetilde{\dot{Q}}|$ overestimate the true heat rate $|\dot{Q}|$. 
Numerically, when the error is $\varepsilon=+10$\,nm~\cite{ritort2006single}, the predicted heat rate is $|\widetilde{\dot{Q}}| = 6.86 \times 10^{-20}$\,W and the true value is $|\dot{Q}|=1.88 \times 10^{-20}$\,W.

\textit{Prediction of energy dissipation rates for active gels.--} \jrg{As another example, we compared the predicted heat rate with experimental measurements for microtuble active gels.} 
In these active materials, kinesin motors cross-link microtubule pairs. 
The forward motion of the motors is driven by the chemical energy from ATP hydrolysis, which also releases energy as heat and sustains self-organized flows at longer length and time scales. 
The first picocalorimetry measurements~\cite{bae2021micromachined} suggest the energy efficiency of these materials is low: the minority of the input chemical energy propagating to productive emergent flows and the majority dissipated away as waste heat~\cite{ActiveGel}.

\jrg{Using the concentrations of chemical species in our main result, Eq.~\eqref{eq:mvspeed}, the predicted heat rate}
\begin{equation}
  \label{eq:activegelexp}
  |\widetilde{\dot{Q}}|= |\partial_{c} \hat{r}| \Delta \hat{c} \Delta \hat{\epsilon} = \frac{|d_{t}{\hat{c}}|}{\Delta \hat{c}} \Delta \hat{\epsilon}
\end{equation}
is in terms of $|d_{t}{\hat{c}}|$ the average rate of ATP hydrolysis and the standard deviations in concentration and energy.
The rate of ATP hydrolysis depends on the initial concentrations of kinesin, microtubule, ATP, and rate constants (SM Sec.~6). 
This rate can be determined with a previously parametrized model or without a model from the numerical derivative of concentration measurements. 
Here, we use measured rate constants and a kinetic model for ATP hydrolysis that were reported previously~\cite{ActiveGel}. 

\jrg{To predict the heat rate with Eq~\eqref{eq:activegelexp}, we can use either measurements or estimates of the concentration and energy fluctuations.
Since there are not direct measurements of the energy fluctuations, we estimate these values from available information.
The active material is in a buffer solution of 1\% 35-kDa polyethylene glycol in room temperature water at a pH of 6.8 (SM Sec.~6). 
A single molecule undergoes about $10^{13}-10^{14}$ collisions per second, exchanging an average energy on the order of $k_BT$ at a rate of $10^{-8}$\,W.} 
\jrg{In Ref.~\cite{ActiveGel}, the heat rate measurements are on the order of $1000$\,s, so we estimate $\Delta\hat{\epsilon} = 10^{-8}$\,W $\times 1000$\,s $ = 10^{-5}$\,J.
This estimate agrees with the uncertainty of the measured heat rate, which are on the order of $100$\,nW with fluctuations are on the order of $10-15$\,nW when the gel was prepared with and without the pyruvate kinase-based ATP regeneration system.}

\jrg{Pipetting error is likely the dominant source of error in these measurements~\cite{ActiveGel}, and data is not currently available for $\Delta\hat{c}$, so we instead estimate the standard deviation using experimental error.}
The total sample volume 
is $0.5$\,$\mu$L and the initial ATP concentration varies from $1.5-1500$\,$\mu$M.
Propagating the 5\% error in dose volume $\Delta \hat{c}=\hat{c}V^{-1} \Delta
V$, the uncertainty in concentration is on the order of 1\,$\mu$M for ATP
concentrations in the range $12.5$-$190$\,$\mu$M. 
Other potential sources of uncertainty are the pipetting protocol and fitted rate constants, which could also be accounted for using Eq.~\eqref{eq:mvspeed}. 

\begin{figure}
\centering
\includegraphics[width=0.75\columnwidth]{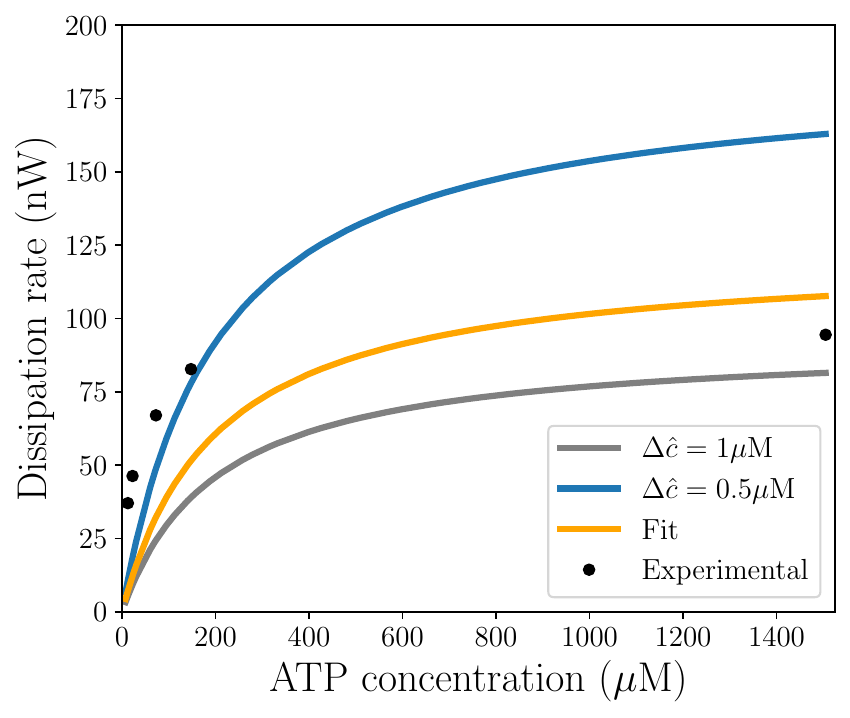}
\caption{\label{fig:3} Comparison of predicted heat rate with experimental measurement. Averaged dissipation rate versus ATP concentration with $[\text{K401}] = 210$\,nM, $[\text{MT}] = 16$\,$\mu$M. 
We use $\Delta\hat{\epsilon} = 10^{-8}$\,W $\times 1000$\,s $ = 10^{-5}$\,J and the rate constants: $k_{\text{D,ATP}} = 96.4$\,$\mu$M, $k_{\text{D,MT}} = 17.4$\,$\mu$M, and $k_{\text{cat}} = 44s^{-1}$. 
The fit (orange) is the chemical kinetics model to the data (black points) from Ref.~\cite{ActiveGel}.}
\end{figure}

While less data is available for these estimates, they illustrate the flexibility of this method and the ability to use information on hand.
With these estimated inputs, the heat rate predicted with Eq.~\eqref{eq:activegelexp} agrees well with picocalorimetry measurements~\cite{ActiveGel}. 
Figure~\ref{fig:3} shows the estimated heat dissipation rate as a function of ATP concentration with all other concentrations fixed. 
Both the measured and estimated dissipation rates (SM Sec.~6) increase with the initial ATP concentration.
Using a standard deviation for concentration $\Delta \hat{c}$ of 0.5 or 1\,$\mu$M, the predicted heat rate is less than a factor of two of the experimental values.
\jrg{For low ATP concentrations, the prediction with $\Delta \hat{c}=0.5$\,$\mu$M agrees better with the measured data than the direct fit with a chemical kinetics model~\cite{ActiveGel}.} 

\textit{Conclusions.--}
Fisher information is widely used in the statistical design of experiments: with known experimental errors, one can predict \textit{a priori} the minimum error in a measured quantity. 
\jrg{By leveraging its appearance in the thermodynamic speed limit, we can use
convenient transformations of the speed limit together with statistical regression hypotheses to predict dissipation rates.
A single speed -- the square root of the Fisher information -- bounds multiple dissipation rates, so we expect the results here to produce estimates of other physical observables.} 
These predictions could be useful in determining the sensitivity of measurements to
experimental control variables and in determining the experimental uncertainty
needed for accurate measurements of dissipation rates.
Used in these ways, the thermodynamic speed limit set by the Fisher information
and the transformation to more convenient variables is a potentially efficient method to guide both the design of experiments and synthetic active materials.
Since these speed limit predictions of dissipation rates can be independent of an underlying microscopic model for the dynamics and the distance from thermal equilibrium, they hold across the length and time scales relevant for active materials.
 
\section{Acknowledgments}
\begin{acknowledgments}
We gratefully acknowledge Peter Foster and Joost Vlassak for sharing the data from Ref.~\cite{ActiveGel}. This material is based upon work supported by the National Science Foundation under Grant No. 2231469. 
\end{acknowledgments}

\bibliography{references} 

\end{document}


\title{Supplemental Material: Dissipation rates from experimental uncertainty}

\author{Aishani Ghosal}
\author{Jason~R.~Green}
\email[]{jason.green@umb.edu}
\affiliation{Department of Chemistry,\
	University of Massachusetts Boston,\
	Boston, MA 02125 USA
}
\affiliation{Department of Physics,\
	University of Massachusetts Boston,\
	Boston, MA 02125 USA
}
\date{\today}

\begin{abstract}	

\end{abstract}	

\maketitle

\section{1. Fisher information from uncertainty}

\jrg{In this section, we show how to use a known or hypothesized functional relationship between $\hat{r}$ and $\hat{x}$ to relate their moments, the Fisher information $\Delta\hat{r}^2$ and $\Delta\hat{x}^2$, without knowing the exact form of their distributions.
To derive our main result in Eq.~5,
\begin{equation}
  \tilde{\tau}^{-2}=\Delta \hat{r}^2 
  = \nabla f^\top(\hat{\boldsymbol{X}})\,\boldsymbol{C}_{\hat{\boldsymbol{X}}}\,\nabla f(\hat{\boldsymbol{X}}),
\end{equation}
and the special case in Eq. 2
\begin{equation}
\tilde{\tau}^{-1}=\Delta\hat{r} = \frac{\Delta \hat{x}}{|\partial_{f}\hat{x}|} + \mathcal{O}(\Delta \hat{x}^2),
\end{equation}
of the main text, we first show how $\Delta \hat{x}$ is related to $\Delta \hat{r}$ using forward error propagation and derive optimal values for $\partial_{f}\hat{x}$ using a linear regression hypothesis.} 

\subsection{Gauss's error propagation law}

To find the speed limit set by the Fisher information, we use a generalization of Gauss's law \cite{Barlow} from error propagation. 
Consider the outcomes of measurements of $n$ observables be represented by random variables $X_i$ with $i=1,\ldots, n$.
Each $X_i$ represents a distinct measurable quantity with its own population mean $\langle X_i\rangle = \mu_{i}$ and variance $\sigma_{i}^2$. 
Direct measurements of $X_i$ can be transformed into indirect measurements of another quantity $Y$. 
For example, if we model $X$ as a random variable with population mean $\mu_X$, population variance $\sigma_{X}^2$, and distribution $p_X(x)$, uncertainties in the measurements of $X$ will propagate to the indirectly measured quantity $Y$.

Suppose we are interested in indirectly measuring the random variable $Y$ and that it is a known function $Y=f(X_1,X_2,\ldots,X_n)$ of the $X_i$.
Using Gauss's error propagation law for uncorrelated $X_i$, we can predict first two moments of the distribution $p_Y(y)$ of $Y$. 
The mean is $\mu_Y=f(\mu_1,\mu_2,\ldots)$ and the variance is
\begin{equation}
	\sigma_{Y}^{2} = \sum_i \left(\frac{\partial f}{\partial X_i}\Big|_{X_i=\mu_{i}}\right)^{2} \sigma_{i}^{2}.
\end{equation}
So, loosely speaking, Gauss's law and its generalization use a known function $f$ between two quantities, $X$ and $Y$, to relate the first two moments of their distributions. 

In practice, one might instead use the realizations $x_i$ of $X_i$ and the estimator $\hat{X}_i$. 
To construct an estimator $\hat{Y}$ of $Y$, we can use an unbiased estimator $\hat X_i$ for $\mu_i$ with realizations $\hat x_i$.
In this case, one can express Gauss's law
\begin{equation}
	\Delta \hat{Y}^{2} =  \sum_i\left(\frac{\partial \hat{Y}}{\partial X_i}\Big|_{X_i=\hat{X}_i}\right)^{2} \Delta \hat{X}^{2}
\end{equation}
in terms of the measured values.
This approximate form can be applied to data. 

\subsection{Forward propagation of errors}

A generalization of Gauss's error propagation law is necessary when the $X_i$ are correlated random variables. 
To derive this result, the quantity $Y$ is again a function of the measured observables $X_i$: 
\begin{equation}\label{suppl}
Y = f(\langle X_1\rangle, \langle X_2\rangle,\ldots, \langle X_n\rangle) =: f(\langle X_i\rangle),
\end{equation}
where $\langle \cdot \rangle$ denotes averaging over the distribution of the variables, $p_{i}(x_i)$.
We are after an estimator for $Y$
\begin{equation}
\hat{Y} = f(\hat X_1, \hat X_2,\ldots, \hat X_n) =: f(\hat X_i)
\end{equation}
in terms of the estimators $\hat{X}_i$. 
We take the estimators of $X_i$ to be unbiased, $\langle \hat{X}_i\rangle=\mu_i$.
However, the mean $\langle\hat{Y}\rangle$ will, in general, differ from $f(\langle X_i\rangle)$.
If the function $f$ is a linear, then:
\begin{equation}
\langle\hat{Y}\rangle = \langle f(\hat X_i)\rangle \stackrel{\text{unbiased}}{=} f(\langle X_i\rangle) = Y,
\end{equation}
and $\hat{Y}$ is an unbiased estimator of $Y$.

Taylor expanding the function $f$ around the mean values $\mu_{i}$ of the measured variables $\hat{X}_{i}$ (for $i=1,2,\ldots,n$), we have 
\begin{equation}
  f \approx f(\mu_{1},\mu_{2},...\mu_{n})+\sum_{i=1}^{n}\left.\frac{\partial f}{\partial \hat{X}_{i}}\right|_{\hat{X}_{i}=\mu_{i}}(\hat{X}_{i}-\mu_{i}).
\end{equation}
Denoting $\partial f/\partial \hat{X}_{i}$ by $f^{\prime}$, $\langle f^{2} \rangle$ becomes
\begin{equation}
	\langle f^{2} \rangle = \left \langle \left(f(\mu_{1},\ldots,\mu_{n})+\sum_{i=1}^{n}f^{\prime}|_{\hat{X}_{i}=\mu_{i}}(\hat{X}_{i}-\mu_{i})\right)^{2} \right \rangle.
\end{equation}
Expanding the right side, we find
\begin{equation}
	\begin{split}
		\langle f^{2} \rangle  & = \left\langle f(\mu_{1}, \ldots, \mu_{n})^{2} \right\rangle \\
		& \quad + \left\langle \sum_{i=1}^{n} \left( f^{\prime}(\hat{X}_{i}) \big|_{\hat{X}_{i}=\mu_{i}} (\hat{X}_{i} - \mu_{i}) \right)^{2} \right\rangle \\
		& \quad + 2 \left\langle \sum_{i=1}^{n} f^{\prime}(\hat{X}_{i}) \big|_{\hat{X}_{i}=\mu_{i}} (\hat{X}_{i} - \mu_{i}) f \right\rangle.
	\end{split}
\end{equation}
Denoting the derivative as $f^{\prime}(\hat{X}_{i})|_{\hat{X}_{i}=\mu_{i}} = f^{\prime}(\mu_{i})$,
the variance as $\langle (\hat{X}_{i}-\mu_{i})^{2} \rangle := \Delta \hat{X}_{i}^{2}$, and the covariance as $\langle(\hat{X}_{i}-\mu_{i})(\hat{X}_{j}-\mu_{j})\rangle := \operatorname{cov}(\hat{X}_{i},\hat{X}_{j})$, we can express $\langle f^{2}\rangle$ as:
\begin{equation}
	\begin{split}
		\langle f^{2}(\mu_{1}, \ldots, \mu_{n}) \rangle & 
		= \left( \sum_{i=1}^{n} f^{\prime}(\hat{X}_{i}) \big|_{\hat{X}_{i}=\mu_{i}} \right)^{2} \Delta \hat{X}_{i}^{2} \\
		& \quad + 2 \sum_{i,j=1}^{n} f^{\prime}(\mu_{i}) f^{\prime}(\mu_{j}) \operatorname{cov}(\hat{X}_{i}, \hat{X}_{j}) \\
		& \quad + 2 \sum_{i=1}^{n} f^{\prime}(\hat{X}_{i}) \big|_{\hat{X}_{i}=\mu_{i}} \\
		& \left\langle f \big|_{\hat{X}_{i}=\mu_{i}} (\hat{X}_{i} - \mu_{i}) \right\rangle.
	\end{split}
\end{equation}
To calculate the variance of $f$, we first find the mean of the function 
\begin{equation}
	\langle f \rangle = \langle f|_{\mu_{1}, \mu_{2}, \dots, \mu_{n}} \rangle + \sum_{i=1}^{n} f^{\prime}(\hat{X}_{i}) \big|_{\hat{X}_{i}=\mu_{i}} \left \langle (\hat{X}_{i} - \mu_{i}) \right \rangle.
\end{equation}
The second term of the above equation vanishes, as $\langle \hat{X}_{i} \rangle = \mu_{i}$. 
Therefore, the variance in $f$ is
\begin{equation}
	\begin{split}
		\Delta f^{2} & =\sum_{i=1,2,...,n}(f^{\prime}(\hat{X}_{i}=\mu_{i}))^{2}\Delta \hat{X}_{i}^{2} \\ &
		+2\sum_{i,j=1,2,...,n}f^{\prime}(\mu_{i})f^{\prime}(\mu_j)\operatorname{cov}(\hat{X}_{i},\hat{X}_{j}).
	\end{split}
\end{equation}
This expression is the generalization of Gauss's error propagation law for a single $f$. 
In vector-matrix form,
\begin{equation}
  \Delta f^2 = \nabla f^\top(\hat{\boldsymbol{X}})\,\boldsymbol{C}_{\hat{\boldsymbol{X}}}\,\nabla f(\hat{\boldsymbol{X}}),
\end{equation}
it depends on the covariance matrix $(\boldsymbol{C}_{\hat{\boldsymbol{X}}})_{ij}=\operatorname{cov}(\hat{X}_{i},\hat{X}_{j})$ and the gradient $\nabla f(\hat{\boldsymbol{X}})$. 
Unlike Gauss's law, it holds even when the measured variables $X_i$ in $\hat{\boldsymbol{X}}=(\hat{X}_1,\hat{X}_2,\ldots)^\top$ are correlated. 

Now consider the function to be the surprisal rate $f=\hat{r}=-\partial_t\ln p$, then the variance of $\hat{r}$ (the Fisher information) becomes
\begin{equation}
  \Delta \hat{r}^2 
  = \nabla f^\top(\hat{\boldsymbol{X}})\,\boldsymbol{C}_{\hat{\boldsymbol{X}}}\,\nabla f(\hat{\boldsymbol{X}}),
  \label{Eq:SMeq6}
\end{equation}
Equation 6 of the main text. 
Equation 2 of main text is a special case.
When the variables $\hat{X}_i$ are uncorrelated, the covariance matrix $\boldsymbol{C}_{\hat{\boldsymbol{X}}}$ is diagonal with elements $\Delta \hat{X}_{i}^2$, reducing this expression to Gauss's formula. 

\subsection{\jrg{Optimal parameters of linear regression hypotheses}}

\jrg{Consider the rate of change of (log) probability, $\hat{r}$ to be a function of the variables, $f(\hat{\boldsymbol{X}})$, where $\hat{\boldsymbol{X}}^{\top}= (\hat{X}_{1},\hat{X}_{2},\cdot,\hat{X}_{N})$. For the function
\begin{equation}
	f(\hat{\boldsymbol{X}}) = a + \boldsymbol{b}^\top \hat{\boldsymbol{X}} +\mathcal{E},
\end{equation}
we can find the optimal parameters by minimizing the mean-squared error:
\begin{equation}
	\langle \mathcal{E}^2 \rangle = \langle (f -(a+\hat{\boldsymbol{X}}^{\top} \boldsymbol{b}))^\top (f -(a+\hat{\boldsymbol{X}}^{\top} \boldsymbol{b})) \rangle.
	\label{eq:MSE}
\end{equation}
Setting $\langle \frac{\partial \mathcal{E}^2}{\partial a} \rangle = \langle \frac{\partial \mathcal{E}^2}{\partial \boldsymbol{b}} \rangle = 0$, 
gives
\begin{equation}
	a = \langle f \rangle -\langle \hat{\boldsymbol{X}}^{\top} \rangle \boldsymbol{b}
	\label{eq:optimalabsol1}
\end{equation}
and
\begin{equation}
	\langle \hat{\boldsymbol{X}} \hat{\boldsymbol{X}}^{\top} \rangle \boldsymbol{b} = \langle \hat{\boldsymbol{X}} (f -a)\rangle.
	\label{eq:optimalabsol2}
\end{equation}
Solving Eq.~\eqref{eq:optimalabsol2}, we find the optimal slope
\begin{equation}
	\boldsymbol{b}_{\text{opt}} = \frac{\operatorname{cov}(\hat{\boldsymbol{X}},f)}{(\langle \hat{\boldsymbol{X}}\hat{\boldsymbol{X}}^{\top})\rangle -\langle \hat{\boldsymbol{X}} \rangle \langle \hat{\boldsymbol{X}}^{\top} \rangle)}.
	\label{eq:optimalb}
\end{equation}
The covariance in the numerator corresponds to physical observables (e.g., heat rate, rate of dissipated work, entropy production rate). 
Substituting $\boldsymbol{b}_{\text{opt}}$ into Eq.~\eqref{eq:optimalabsol1}, we find the optimal intercept $a_{\text{opt}}$.}

\jrg{In the next section, we consider a Brownian particle being pulled by a harmonic trap. There, we take the driven particle position to have the estimator $\hat{x}$. 
From Eq.~\eqref{eq:optimalb}, the optimal slope is $\boldsymbol{b}_{\text{opt}} = \langle d_{t}\hat{x} \rangle/\Delta \hat{x}^{2} = v/\Delta \hat{x}^{2}$.
Using this slope, Eq.~\eqref{Eq:SMeq6} becomes
\begin{equation}
	\label{eq:mainerrorpropagationformulaonevar}
	\Delta\hat{r} = \frac{\Delta \hat{x}}{|\partial_{f}\hat{x}|} = \frac{|v|}{\Delta \hat{x}},
\end{equation}
Eq. 2 of the main text.
As we describe in the main text, this is a coordinate-transformation of the speed limit. 
Measuring the variance of the particle position $\Delta \hat{x}^2$ while pulling the trap at a speed $v$ is all that one needs to get the optimal slope.}

\section{2. Pulled Brownian particle}

From the speed limit, we can calculate the heat rate for the dragged Brownian particle from position measurements.
For overdamped Langevin dynamics, we compare our prediction to the mean heat rate from an analytical formula~\cite{Imparato} and our numerical simulations.
We expand on these results below and discuss how the heat rate expression is independent of the reference frame.

\subsection{Heat rate from coordinate-transformed speed limit}

\jrg{The Fisher information depends on the chosen parametrization. Thus, we can consider alternatives that are more experimentally accessible.} 
As an example, we consider measurements of the position of a colloidal particle being pulled through a viscous medium. 
Building on the previous section, we use $\hat{x}$ to represent the position estimator and take the estimator of the rate $r$ to be a function of the data $\hat{r}=f(\hat{x})$. 
The main insight here is that the uncertainty $\Delta \hat{x}^2$ will propagate to the Fisher information $\Delta\hat{r}^2$. 
Said differently, one can reparametrize the Fisher information $I_F=\tau^{-1}$ in terms of the position as $\tilde{I}_F=\tilde{\tau}^{-1}$.
To transform the Fisher information from time to position coordinates, we use
\begin{equation}
	\label{eq:jacobian}
	\frac{\partial f}{\partial \hat{x}}\Big|_{\hat{x}=vt} = \frac{ d_{t}\langle x \rangle}{\Delta \hat{x}^2}
	= \frac{v}{\Delta \hat{x}^2},
\end{equation}
which follows from Eq.~\eqref{eq:optimalb}.
To obtain the optimal slope, one only needs to measure the variance of the particle position $\Delta x^2$ while pulling the trap at a speed $v$. Together with Eq.~6 of the main text,  
the Fisher information parametrized by position becomes
\begin{equation}
\label{eq:EPFisherinformation}
\Delta\hat{r}^2 =\tilde{\tau}^{-2} = \tilde{I}_F = \left(\frac{\partial f}{\partial \hat{x}}\Big|_{\hat{x}=v t}\right)^2 \Delta \hat{x} ^{2} = \frac{v^2}{\Delta \hat{x}^2}.
\end{equation}
This transformation of the Fisher information only depends only on the speed of the harmonic trap $v$, which is experimentally controlled, and the uncertainty in the particle position $\Delta \hat{x}$.

The linear relation between $x$ and $\hat{r}$ is a sufficient condition for the saturation of the thermodynamic speed limit \cite{Nicholson:2020}, $\tilde{\tau}_{Q}^{-1} = \tilde{\tau}^{-1}$.
Physically, saturation means that incorporating the measurement error makes the time for $Q$ to evolve by $\Delta \hat{\epsilon}$ equal to the time it would take for a probability distribution to change to a distinguishable state.
From the saturated speed limit, the predicted heat becomes
\begin{equation}
	\label{eq:predictedheat}
	|\widetilde{\dot{Q}}| = \sqrt{\tilde{I}_F}\Delta\hat{\epsilon} = \frac{|v| \Delta\hat{\epsilon}}{\Delta \hat{x}},
\end{equation}
Eq.~(3) in the main text. 

Next, we assume an overdamped Langevin dynamics and compare the predicted heat rate to the heat rate from an analytical formula and numerical simulations.

\subsection{\jrg{Overdamped Langevin dynamics}}

We analyzed the overdamped motion of a Brownian particle subjected to a time-dependent conservative force.
In the laboratory frame, the particle position varies over time
\begin{equation}
	\label{eq:LDymx}
	\frac{dx}{dt} = -\gamma\frac{\partial}{\partial x} U(x(t),\lambda(t)) + \xi(t)
\end{equation}
due to both random thermal fluctuations from a thermal bath at temperature $T$ and deterministic forces modified by the time dependent control parameter $\lambda$.
Here, the particle is trapped by a harmonic potential $U = \frac{1}{2}k_{f}(\hat{x}-vt)^2$ with a stiffness $k_f$ and the trap is translated at constant speed $|v|$ for $t>0$. 
At times $t>0$, the particle position will deviate on average from the trap center.
According to Stokes' law, a spherical particle of radius $r$ in a fluid with viscosity $\eta$ has a friction coefficient given by $6\pi\eta r$.
The random force $\xi(t)$ is white Gaussian noise with zero mean and delta function correlation, $\left\langle  \xi(t)  \xi(t^{\prime})\right\rangle = 2 \gamma k_BT \delta(t-t^{\prime})$.
Particle motion is damped by viscous drag $-\gamma^{-1}d_{t}{\hat{x}}$, where $\gamma$ is the inverse friction coefficient.
The time-dependent conservative force does work in pulling the particle which, on average, is dissipated as heat.

First, the average heat rate can be calculated analytically from the distribution of heat exchanged with the environment at time $t$.
As shown in Ref.~\cite{Imparato}, the distribution is
\begin{equation}
	P(Q, t \gg \mathcal{T}) = \sqrt{\frac{\beta\gamma}{4 \pi v^{2} t}} e^{-\frac{\beta\gamma}{4v^{2}t} \left(Q+\frac{t v^2}{\gamma}\right)^2}
\end{equation}
with $\beta=1/{k_{B}T}$, the relaxation time of the particle $\mathcal{T} = (\gamma k_{f})^{-1}$. 
In this limit, the average dissipation is
\begin{equation}
	|\langle Q(t>\mathcal{T}) \rangle| = \int Q P(Q, t) dQ = \frac{v^{2}t}{\gamma}.
\end{equation}
Experimental values used from Ref.~\cite{Imparato} are:
velocity, $v=10^{-6}$\,m/s; force constant, $k_{f}=6.67 \times 10^{-7}$\,Nm$^{-1}$; viscosity of water at 296.5K, $\eta=10^{-3}$\,Pa s; radius of the particle, $r=1\mu$\,m; and inverse friction coefficient, $\gamma =(6\pi\eta r)^{-1} = 5.30 \times 10^{7}$\,mN$^{-1}$s$^{-1}$. 
Substituting the parameter values, the dissipation rate is $|\langle\dot{Q}\rangle| = v^{2}/\gamma = 1.885 \times 10^{-20}$\,W. 

For short times $t < \mathcal{T}$, the dissipated energy rate during the time interval $t$ and $t+\Delta T$ from Ref.~\cite{Vanzon} is
\begin{equation}
	|\langle \dot{Q}(t < \mathcal{T}) \rangle| =  \frac{v^{2}(e^{-2 t-2 \Delta T }+1)}{\gamma},
\end{equation}
which is $3.71429 \times 10^{-20}$\,W for $t=0.01$\,s and $\Delta T=0.005$\,s.

As another confirmation of our predicted heat rate, we simulated trajectories from the Langevin dynamics in Eq.~\eqref{eq:LDymx}. 
For a single noise realization, the heat rate $\dot{Q} = \partial_{x} U \circ \dot{x} = -F(x) \circ \dot{x}$ has the discrete representation
\begin{equation}
	\dot{Q} = -\frac{1}{2n \Delta t}\sum_{i=1}^{n} (F(x_{i-1})+F(x_{i}))(x_i-x_{i-1}),
	\label{eq:heatratesimE}
\end{equation}
where $n$ is the number of simulation steps, $\circ$ is the Stratonovich product, and $\Delta t$ is the sampling time. 
The histogram for the heat rate shown in Figure 1 of the main text is for 10000 noise realizations over a total time duration of $4$\,s using the parameter values above and a sampling time $\Delta t= 1$\,ms. 
The absolute value of the mean heat rate from histogram is $1.85 \times 10^{-20}$ J s$^{-1}$.
Figure 1 of main text also compares our predicted heat rate, the analytical mean heat rate and the value from our simulations.

\subsection{\jrg{Frame invariance of heat rate}}

\jrg{According to Sekimoto ~\cite{Sekimoto}, the heat transferred by a particle to its surroundings along a particular stochastic trajectory connecting two microstates can be identified with the change in its internal energy $U(x(t),\lambda(t))$ as a result of changes in its coordinates $x(t)$ at a fixed value of a control parameter $\lambda$. 
Over an interval of time $t$, therefore, the total amount of heat $Q(t)$ dissipated into the medium during a single stochastic realization is  
\begin{equation}
	Q(t)=\int^{t}_{0} dt^\prime \dot{x}(t^\prime)  \circ \frac{\partial U(x(t^\prime), \lambda(t^\prime))}{\partial x}.
	\label{eq:heat}
\end{equation}
However, Sekimoto's definitions of work and heat are not frame-invariant~\cite{Speck,Sabine} as the explicit time dependence of potential energy vanishes in a comoving frame.}

\jrg{For a trap velocity $v$, conservative force $F$, and nonconservative force $f$, the rate of work becomes
\begin{equation}
	\dot{W}(t) = \partial_{t}U(x,t) - vF(x,t) + f(x,t)(\dot{x} - v).
	\label{eq:workdef1}
\end{equation}
The rate of heat transfer can then be derived from energy conservation
\begin{equation}
	\dot{Q}(t) = \dot{W}(t) - \frac{dU}{dt},
	\label{eq:heatdef1}
\end{equation}
where the total derivative $d_tU$ consists of two terms: $d_tU = \partial_{t}U + \dot{x} F$.
Substituting Eq.~\ref{eq:workdef1} into Eq.~\ref{eq:heatdef1}, we find
\begin{equation}
	\dot{Q}(t) = (f + F)  \circ (\dot{x} - v).
	\label{eq:fiheatdef1}
\end{equation}
In the co-moving frame $\dot{x} = v$, which makes $\dot{Q}$ vanish.
In contrast, for the lab frame where $\dot{x} = 0$ and $f=0$ in our example, the heat rate is $\dot{Q} = -F  \circ v$ (Eq.~\eqref{eq:heatratesimE}).
Note that we use the convention that heat dissipated to the surrounding $Q < 0$.}

\bigskip

\section{3. Heat rate from coordinate-transformed speed limit}

\jrg{In this section, we discuss the heat rate predicted by the coordinate-transformed speed limit and apply our approach to a system that is only partially observed and the effect of missing information about the driven degrees of freedom on the predicted dissipation.}

\subsection{\jrg{Heat rates and speed limit do not vanish at nonequilibrium steady state}}

\jrg{Eq.~3 in the main text corresponds to the housekeeping heat rate in the laboratory frame.}

\sda{The total dissipated heat decomposes into the housekeeping heat $Q_{hk}$, which is the heat exchanged needed to maintain a nonequilibrium steady state, and the excess heat $Q_{ex}$, which is the heat exchanged during transient changes in nonequilibrium state (e.g., relaxation). 
This decomposition traces back to ref.~\cite{Oono}. 
For any stochastic process relaxing to equilibrium, $Q_{hk} = Q_{ex} = 0$. At a nonequilibrium steady state, $\langle Q_{ex} \rangle = 0$ and $\langle Q_{hk} \rangle \leq 0$.}

\sda{The heat constrained by the time-parameterized speed limit is derived from a total derivative of the time-independent potential.
Therefore, at long time when the system reaches the equilibrium steady state, the housekeeping heat vanishes.
The same logic applies to our example of a Brownian particle dragged by a moving harmonic trap in the comoving frame, where the particle is pulled at a constant velocity.
As a result, over time, it reaches an equilibrium steady state, and the housekeeping heat rate becomes zero.
However, in the lab-fixed frame, the particle evolves under a time-dependent conservative potential, reaching a non-equilibrium steady state and releasing heat into the heat bath with which it is in contact.
This heat is released steadily as non-zero housekeeping heat.
In the second example of active gel consisting of microtubule, kinesin motors and ATP molecules also reaches to a non-equilibrium steady state and dissipates housekeeping heat.}

\subsection{\jrg{Heat rate estimates for partially observed systems}}

\jrg{Measuring fluctuations in the dissipative variables is important for accurate heat rate predictions using our approach.
For example, suppose the pulled colloidal particle is confined to a trap moving along $x$-coordinate and the particle is static in the $y$ coordinate.
Take $x$ to be the dissipative variable and imagine we measure fluctuations in another variable $y$.
Assuming $y$ is correlated with $x$, predicting the heat rate with its fluctuations will be a less accurate when $y$ is driven and especially poor when $y$ is in equilibrium.
If the $y$ coordinate of the particle is equilibrated and does not contribute to the dissipation, then using $y$ as the sole coordinate for estimating the heat rate would provide a poor estimate. 
However, if the equilibrated $y$ coordinate is coupled to the driven $x$ coordinate, the estimate would be trivial.}

\jrg{In this two-dimensional case, we can consider a linear regression hypothesis for the surprisal rate $\hat{r}(x,y) = ax + by +c$ and find the variance $\tilde{\tau}^{-2} \equiv \tilde{I}_{F} \equiv \Delta \hat{r}^{2}$ representing the coordinate-transformed Fisher information. 
In terms of the variance in position, this Fisher information is
\begin{equation}
	\begin{split}
		\tilde{\tau}^{-2} &= \left( \frac{\partial \hat{r}}{\partial x} \right)^{2} \Delta x^{2} + \left( \frac{\partial \hat{r}}{\partial y} \right)^{2} \Delta y^{2} \\
		& + 2 \left( \frac{\partial \hat{r}}{\partial x} \right) \left( \frac{\partial \hat{r}}{\partial y} \right) \operatorname{cov}(x,y).
	\end{split}
\end{equation}
In this scenario, we do not have access to the driven $x$ coordinate, so terms involving $(\partial \hat{r}/\partial x)$ vanish. 
The optimal values for the linear statistical model yield $(\partial \hat{r}/\partial y) = \partial_{t} y/\sigma_{y}^{2}$. 
Since $y$ is equilibrated, $\partial_{t} y$ vanishes, leading to a poor estimate of the heat rate when using the $y$ coordinate for the coordinate transformation.}

\bigskip

\section{4. Validation of Eq. 7 in the main text}

In this section, we demonstrate that Eq. 7 of \jrg{the main text} also holds even when $\hat{r}$ is a nonlinear function of $\hat{x}$. 
We begin with a general function $f(x)$ and then apply the result to a model of Brownian motion.

We start with the Taylor expansion of the function around its mean value, $\langle \hat{x} \rangle =  0$:
\begin{equation}
	\label{eq:Taylorexp}
	f \approx f(\mu) + f^{(1)}(\mu)(\hat{x}-\mu) + \frac{1}{2!} f^{(2)}(\mu) (\hat{x}-\mu)^{2} + ....
\end{equation}
Here, $f^{(n)}(\langle \hat{x} \rangle)$ represents the $n$th-order partial derivative of $f$ with respect to the variable $x$, evaluated at the mean $\langle\hat{x}\rangle$. 
For $f=\hat{r}$, the first derivative $f^{(1)}(\langle \hat{x} \rangle)$ vanishes. 
At $f^{\prime}(\langle \hat{x} \rangle)=0$, this equation simplifies to $f \approx f(0) + \frac{1}{2!} f^{\prime\prime}(0) \hat{x}^{2} +\ldots$. 

The variance of $f$ can be computed from the first moment, $\langle f \rangle$, and the second moment, $\langle f^{2} \rangle$:
\begin{equation} 
	\label{eq:fm} 
	\langle f \rangle ^{2} = f^{2}(0) + \left(\frac{f^{\prime\prime}(0)}{2!}\right)^{2} \langle \hat{x}^{2} \rangle^{2} + f(0) f^{\prime\prime}(0) \langle \hat{x}^{2} \rangle, 
\end{equation} 
and 
\begin{equation} 
	\label{eq:sm}
	\langle f^{2} \rangle = f^{2}(0) + \left(\frac{f^{\prime\prime}(0)}{2!}\right)^{2} \langle \hat{x}^{4} \rangle + f(0) f^{\prime\prime}(0) \langle \hat{x}^{2} \rangle. 
\end{equation}
Subtracting Eq. \eqref{eq:sm} from Eq. \eqref{eq:fm}, we obtain: 
\begin{equation} 
	\langle f^{2} \rangle - \langle f \rangle ^{2} = \left(\frac{f^{\prime\prime}(0)}{2!}\right)^{2} \left(\langle \hat{x}^{4} \rangle - \langle \hat{x}^{2} \rangle^{2}\right). 
\end{equation}
As $f = \hat{r}$, we find the variance in $\hat{r}$, relating it to the squared inverse intrinsic timescale:
\begin{equation}
	\label{eq:mainVariance}
	\tilde{\tau}^{-2} := \Delta\hat{r}^{2} = \left(\frac{\partial^{2}\hat{r}(\langle \hat{x} \rangle)}{2!\partial \hat{x}^2}\right)^{2} (\Delta \hat{x}^{2})^{2}.
\end{equation} 
This is the coordinate-transformed Fisher information. From this expression for the intrinsic timescale, we apply it to a model system that follows a quadratic relation between $\hat{x}$ and $\hat{r}$.

Consider a free Brownian particle diffusing in a medium with diffusion coefficient $D$. 
The probability distribution of position $x$ at time $t$ is 
\begin{equation}
	P(x,t)=\frac{e^{-\frac{x^2}{4 D t}}}{\sqrt{4 \pi Dt}}.
\end{equation}
Differentiating $\hat{r}$ with respect to $x$, we have
\begin{equation}
	\label{eq:errorexpinI}
	\frac{\partial \hat{r}}{\partial x} = \frac{1}{P}\frac{\partial^{2}J}{\partial x^2}-\frac{1}{P^2}\frac{\partial J}{\partial x} \frac{\partial P}{\partial x}.
\end{equation}
We can rewrite this expression using $J=D\partial P(x,t)/\partial x$, which is the solution to the diffusion equation $\partial P(x,t)/\partial t = D \partial^{2} P(x,t)/\partial x^2$, and recognize that $\partial^{n} P(x,t)/\partial x^{n}|_{\langle \hat{x} \rangle=0} = 0$ for odd $n$.
Then, differentiating both sides of Eq.~\eqref{eq:errorexpinI} gives
\begin{equation}
	\begin{split}
		\label{eq:Einx}
		\tilde{\tau}^{-2} & =  \frac{1}{4} \left(\frac{\partial^{2}\hat{r}}{\partial x^2}\right)^{2} (\Delta \hat{x}^{2})^2 \\&
		= \frac{1}{4} \Big(\frac{DP^{(4)}}{P}|_{\hat{x}=\langle \hat{x} \rangle}-\frac{D(P^{(2)})^{2}}{P^{2}}|_{x=\langle \hat{x} \rangle}\Big)^{2}(\Delta \hat{x}^{2})^2.
	\end{split}
\end{equation}
After simplifying, we find 
\begin{equation}
	\label{eq:EPFisherfromx}
	\tilde{\tau}^{-2} = \frac{1}{4} \left(\frac{1}{2Dt^{2}}\right)^{2} (\Delta \hat{x}^2)^2.
\end{equation}
The Fisher information in this case is
\begin{equation}
	I_{F} = \tau^{-2} 
	= \int_{-\infty}^{\infty}  P(x,t) \left(\frac{\dot{P}}{P}\right)^{2}\, dx
	=\frac{t^{-2}}{2}.
\end{equation}
The ratio of the squared intrinsic timescale to the estimate is
\begin{equation}
\label{eq:mainsecondresult}
\frac{\tilde{\tau}^{-2}}{\tau^{-2}} =
\frac{(\Delta  \hat{x})^2}{\sigma_{x}^2}.
\end{equation}
The right hand side has the population parameter $\sigma_{x} = 2Dt$. 
Thus, Eq.~\ref{eq:mainsecondresult} (Eq. 7 of the main text) is valid when $\hat{r}$ is a nonlinear function of $\hat{x}$.\\

\section{5. Master equation modeling of an active gel}

Now we analytically validate Eq.~7 from the main text for an active gel.
The ATP hydrolysis-driven active gel system described in the main text consumes ATP through a forward reaction with rate constant $k_3$ and produces them through a reverse reaction with rate constant $k_4$.
We model this reaction as a birth-death process with time-independent rate constants.
Let $P_{n}(t)$ denote the probability of having $n$ ATP molecules in the system at time $t$.
The birth rate constant $k_4$ increases the number from $n-1$ to $n$, while the death rate constant $k_3$ decreases it from $n+1$ to $n$.
This results in the ordinary differential equation
\begin{equation}
	d_{t}P_{n}(t) = -(nk_3 + k_4)P_{n}(t) + k_4 P_{n-1}(t) + nk_3 P_{n+1}(t)
\end{equation}
where $k_3$ and $k_4$ are expressed in terms of the equilibrium rate constant $k_{\text{D,ATP}} \equiv k_4/k_3$. The solution of this differential equation is~\cite{Ross}
\begin{equation}
	P_{n}(t) = \frac{1}{n!} e^{-\frac{k_4 \left(1 - e^{-k_3 t}\right)}{k_3}} \left(\frac{k_4 \left(1 - e^{-k_3 t}\right)}{k_3}\right)^n.
\end{equation} 
The distribution has equal mean $\langle n \rangle$ and variance $\sigma_{n}^2$, where $\langle n \rangle = \sigma_{n}^2 = k_4 \left(1 - e^{-k_3 t}\right)/k_3$.

For this continuous-time, discrete-state model, the Fisher information is
\begin{equation}
	I_{F}(t) 
	= \sum_{n=0}^{\infty} \frac{\dot{P}_{n}^{2}(t)}{P_{n}(t)} - \left(\sum_{n=0}^{\infty} \dot{P}_{n}(t)\right)^2.
\end{equation}
$\dot{I} = -d_{t}\ln P_{n}(t)$. The second term on the right-hand side vanishes due to the conservation of probability, and the first term becomes:
\begin{equation}
	I_{F}(t) \equiv \tau^{-2} = \sum_{n=0}^{\infty} \frac{\dot{P}_{n}^{2}(t)}{P_{n}(t)} = \frac{k_4 k_3 e^{-k_3 t}}{e^{k_3 t} - 1}.
\end{equation}
The coordinate-transformed Fisher information is
\begin{equation}
\begin{aligned}
	\tilde{I}_{F} &= \tilde{\tau}^{-2} = \left(\underset{\delta n \to 0}{\text{lim}} \frac{r(n + \delta n) - r(n - \delta n)}{2 \delta n}\right)^{2} \Delta n^{2}\\
	&= \left(\frac{k_3}{e^{k_3 t} - 1}\right)^{2} \Delta n^{2}.
\end{aligned}
\end{equation}
When $\Delta n = \sigma_{n}$ and $\mathcal{E} = 0$, we find $\tilde{I}_{F} = I_{F}(t)$, which confirms Eq.~6 of the main text.

Although the average dissipation rate for a Markov model is known~\cite{Zia_2006,Schnakenberg}
\begin{equation}
	\label{eq:EPRfromME}
	\dot{\Sigma}(t) = \sum_{n} \left(k_{4} P_{n-1}(t) - k_{3} P_{n+1}(t)\right) \ln \left(\frac{k_{4} P_{n-1}(t)}{k_{3} P_{n+1}(t)}\right),
\end{equation}
we cannot currently use Eq.~\eqref{eq:EPRfromME} to estimate dissipation for active gel. The values of $k_{3}$ and $k_{4}$ (the reaction schemes are shown in SM Sec.~6) are experimentally unknown.\\

\section{6. Heat rate prediction for active gel}

To estimate the dissipation rate for active gel using Eq. 8 of main text, we need $d_{t}\hat{c}$. Since these data are not currently available, we obtain time derivative of ATP concentration from a chemical kinetics model. This section derives the expression for $d_{t}\hat{c}$. We relate $d_{t}\hat{c}$ to the rate of the ATP hydrolysis reaction, which occurs in two steps~\cite{Foster}.
First, kinesin binds to microtubules in a first-order reaction with a dissociation constant $K_{\text{D,MT}} \equiv \frac{k_2}{k_1}$:
\[
\underset{\text{Kinesin}}{[\text{K401}]} + \underset{\text{Microtubule}}{[\text{MT}]}
\xrightleftharpoons[k_{2}]{k_{1}}
\underset{}{[\text{K401} \cdot \text{MT}]}.
\]
Next, the ATP hydrolysis reaction starts with an equilibrium binding reaction, characterized by the dissociation constant $K_{\text{D,ATP}} \equiv k_4/k_3$:
\[
[\text{K401} \cdot \text{MT}] + [\text{ATP}] \xrightleftharpoons[k_{4}]{k_{3}}
[\text{K401} \cdot \text{MT} \cdot \text{ATP}],
\]
followed by the hydrolysis reaction with a rate constant $k_{\text{cat}}$:
\[
[\text{K401} \cdot \text{MT} \cdot \text{ATP}]\xrightarrow[]{k_{\text{cat}}} [\text{MT} \cdot \text{K401}] + [\text{ADP}] + \text{P}_{i}.
\]
For these reactions, the ATP hydrolysis rate is
\begin{equation}
	|\langle d_{t}{\hat{c}}\rangle| \equiv \left|d_{t}[\text{ATP}]\right| = -k_{\text{cat}} [\text{K401} \cdot \text{MT} \cdot \text{ATP}].
\end{equation}
Considering the conservation of microtubule and kinesin concentrations, the instantaneous reaction rate can be expressed in terms of the equilibrium rate constants ($k_{\text{D,MT}}$ and $k_{\text{D,ATP}}$) and the initial concentrations of kinesin ($[\text{K401}]_{0}$) and microtubules ($[\text{MT}]_{0}$) \cite{Foster}:
\begin{widetext}
	\begin{equation}
		\label{eq:rateeq}
		\begin{split}
			|\langle d_{t}{\hat{c}}\rangle| \equiv \left|d_{t}[\text{ATP}]\right| &= \frac{ k_{\text{cat}}[\text{ATP}]}{2 ([\text{ATP}]+k_{\text{D,ATP}})} \left([\text{K401}]_{0}+k_{\text{D,MT}}+[\text{MT}]_{0}-k_{\text{D,MT}}\frac{[\text{ATP}]}{[\text{ATP}]+k_{\text{D,ATP}}} \right. \\ 
			& \left. \quad -\sqrt{-4 [\text{K401}]_{0}[\text{MT}]_{0}\left([\text{K401}]_{0}+k_{\text{D,MT}}+[\text{MT}]_{0}- k_{\text{D,MT}}\frac{[\text{ATP}]}{[\text{ATP}]+k_{\text{D,ATP}}}\right)^2}\right).
		\end{split}
	\end{equation}
\end{widetext}
We use this expression as $|d_{t}{\hat{c}}|$ in the main text to calculate the heat rate $|\widetilde{\dot{Q}}|$ (Eq.~8 in the main text).
All parameters are experimentally accessible: $k_{\text{D,MT}}=17.4\,\mu\text{M}$, $k_{\text{D,ATP}}=96.4\,\mu\text{M}$, and $k_{\text{cat}} =44\,\text{s}^{-1}$.

\jrg{In the main text, we also plot $|\widetilde{\dot{Q}}|$ versus $[\text{ATP}]$ (with $[\text{MT}]_{0} = 16$\, $\mu$M and $[\text{K401}]_{0} = 210$\,nM) in Figure 3. 
The fit refers to $\dot{Q} \approx V_{\text{system}} (\Delta H_{\text{K401}} + \Delta H_{\text{PK}}) r_{\text{kinesin}}$ (Eq. 19 of SM~\cite{Foster}).
There, we consider the energy fluctuation as $\Delta \hat{\epsilon} = 10^{-8}$\,J, and we choose two different values for the uncertainty in ATP concentrations: $\sigma_{\hat{c}} = \Delta \hat{c} = 1\,\mu$M and $0.5\,\mu$M. 
We estimated the energy fluctuations to be $10^{-8}$\,J or 10 nW in the active gel based on Foster et al.'s~\cite{Foster} experimental measurements of the heat release rate.
Over 7 minute intervals for a total of 180 minutes, their measurements show fluctuations on the order of 10-15 nW when the gel was prepared with and without the pyruvate kinase-based ATP regeneration system.
As a point of comparison, these fluctuations are smaller than the mean heat release rate they measured: the microtubule active gel dissipates energy at a rate on the order of 100 nW.
The experimental uncertainty is 0.2 nW energy resolution of the picocalorimeter~\cite{Foster}.
Since 10 nW is an order of magnitude smaller than the measured heat release and two orders of magnitude larger than the experimental error, we used the 10 nW deviations as an estimate of the energy fluctuations.}
The active material is formed in a buffer solution of 1\% 35-kDa polyethylene glycol in water at a pH of 6.8.
So, as another estimate, we assumed that a single molecule in room temperature water undergoes about\ $10^{13}-10^{14}$ collisions per second.
The average energy exchanged in each of these collisions is about $k_BT$ (at 298K, $k_BT=4.11\times  10^{-21}$ J).
So, at room temperature, molecules in solution experience random thermal noise that is approximately  $10^{-8}$ W or 10 nW.
This value agrees with the experimental measurements and also supports the use of 10 nW in this example.